\documentclass[reprint,author-numerical,nofootinbib]{revtex4-1}

\usepackage{graphicx}
\usepackage{dcolumn}
\usepackage{bm}
\newtheorem{lemma}{Lemma}
\newtheorem{proposition}{Proposition}
\newtheorem{theorem}{Theorem}
\newtheorem{corollary}{Corollary}

\def\dif{{\rm d}}

\newcommand{\be}{\begin{equation}}
\newcommand{\ee}{\end{equation}}

\begin{document}

\preprint{AIP/123-QED}

\title[A hydrodynamic approach to the classical ideal gas]{A hydrodynamic approach to the classical ideal gas}

\author{Bartolom\'e Coll}
\affiliation{ Departament d'Astronomia i Astrof\'{\i}sica,
Universitat de Val\`encia, E-46100 Burjassot, Val\`encia, Spain.}
\author{Joan Josep Ferrando}
\altaffiliation[Also at ]{Observatori Astron\`omic, Universitat
de Val\`encia,  E-46980 Paterna, Val\`encia, Spain}
\email{joan.ferrando@uv.es.} \affiliation{ Departament d'Astronomia
i Astrof\'{\i}sica, Universitat
de Val\`encia, E-46100 Burjassot, Val\`encia, Spain.}
\author{Juan Antonio S\'aez}
\affiliation{ Departament de Matem\`atiques per a l'Economia i
l'Empresa,
Universitat de Val\`encia, E-46071 Val\`encia, Spain.
}%

\date{\today}

\begin{abstract}
The necessary and sufficient condition for a conservative perfect
fluid energy tensor to
 be the energetic evolution of a classical ideal gas is obtained. This condition forces the square of the speed of sound to have the form $c_s^2 = \frac{\gamma p}{\rho+p}$ in terms of the hydrodynamic quantities, energy density $\rho$ and pressure $p$, $\gamma$ being  the (constant) adiabatic index. The {\em inverse problem} for this case is also solved, that is, the determination of  all the fluids whose evolutions are represented by a conservative energy tensor endowed with the above expression of  $c^2_s$, and it shows that these fluids are, and only are, those fulfilling a Poisson  law. The relativistic compressibility conditions for the classical ideal gases and the Poisson gases are analyzed in depth and the values for the adiabatic index $\gamma$ for which the compressibility conditions hold in physically relevant ranges of the hydrodynamic quantities $\rho, p$ are obtained. Some scenarios that model isothermal or isentropic evolutions of a classical ideal gas are revisited, and preliminary results are presented in applying our hydrodynamic approach to looking for perfect fluid solutions that model the evolution of a classical ideal gas or of a Poisson gas.

\end{abstract}

\maketitle

\section{\label{sec-intro}Introduction}

In the relativistic framework, a perfect fluid is usually assumed to
be a {\em perfect energy tensor}, $T \equiv (\rho+ p) u \otimes u +
p \, g$, solution to the conservation equations $\nabla \cdot T =
0$. It can model a test fluid in any given space-time, or the source
of a solution of the Einstein field equations $G = k T$. Nevertheless, complementary physical
requirements on the {\em hydrodynamic quantities} (the {\em unit
velocity} $u$, the {\em energy density} $\rho$, and the {\em
pressure} $p$) must be imposed for $T$ to represent the energetic
evolution of a realistic thermodynamic fluid in local thermal equilibrium.

As it is well known, a fluid whose all possible energetic evolutions in local thermal equilibrium are described by perfect energy tensors is necessarily Pascalian and with vanishing heat conductivity. Its equations are those of the Eckart's thermodynamic scheme \cite{Eckart} for this case, and lead to the introduction  of the {\em thermodynamic quantities}: {\em matter density} $n$, {\em specific internal energy} $\epsilon$, {\em temperature} $\Theta$, and {\em specific entropy} $s$. On the other hand, we will restrict the fluids obeying these equations to verify general constraints of physical reality, namely the positivity conditions of some of their quantities, the energy conditions \cite{Plebanski} and the relativistic compressibility conditions \cite{Israel}\cite{Lichnero-1}.

Elsewhere \cite{Coll-Ferrando-termo} we have shown that, given a perfect energy tensor in a space-time domain, the question of whether or not it admits the above thermodynamic scheme can be detected by conditions just involving the hydrodynamic quantities $(u, \rho, p)$, namely that the {\em indicatrix function} $\chi = \dot{p}/\dot{\rho}$ be an equation of state, $\chi = \chi(\rho,p)$, and then it is the square of the speed of sound $c^2_s$. This result offers a purely hydrodynamic characterization of the local thermal equilibrium and solves the {\em generic direct problem} \cite{CFS-LTE}, i.e. the  determination of the set ${\bf T}_{\bf F}$ of all perfect energy tensors corresponding to all possible evolutions of the set ${\bf F}$ of all perfect fluids. In \cite{CFS-LTE} we have also solved the {\em generic  inverse
problem} for a perfect energy tensor $T$, i. e. the determination
of the set ${\bf F}_T$ of all perfect fluids for which $T$ is the
energetic description of a particular evolution.

Nevertheless, because of its practical applications, solving a {\em
specific direct problem} for a family of fluids may be more
interesting than to solve the generic one. We have
studied in \cite{CFS-LTE} the specific direct problem for the family 
of the {\em generic ideal gases}, i.e. those defined by the equation of state $p=k n \Theta$. These results have allowed us the study of the Stephani
universes that can be interpreted as an ideal gas in local thermal
equilibrium \cite{C-F}, and the determination of the associated thermodynamics.

Furthermore, elsewhere \cite{CFS-CC} we have shown that two of the
relativistic compressibility conditions can be formulated in terms
of the hydrodynamic quantity $\chi(\rho,p)$, the square of the speed of sound, so that we
have a hydrodynamic characterization for these two constraints.

Here we apply all these hydrodynamic approaches to the {\em
classical ideal gas}, defined by the equations of state, $p=k n \Theta$ and $\epsilon
= c_v \Theta$. The interest of this study is twofold. On one hand,
the classical ideal gas is defined by simple equations of state and
it allows us to go further away in obtaining and interpreting the
results, so that the present study is useful to gain a better
understanding of the concepts and conclusions obtained in
\cite{CFS-LTE} and \cite{CFS-CC}. On the other hand, the classical
ideal gas is a good model for realistic fluids in the low
temperature range, so that our results can be applied in looking for new
physically interesting solutions, or in interpreting already known
ones in this range.

In Section \ref{sec-lte} we present the basic elements concerning
the local thermal equilibrium assumption, and we outline some
results on the above-mentioned hydrodynamical approach.

Section \ref{sec-CIG} is devoted to studying the specific direct and
inverse problems for the classical ideal gas (CIG): we show that the
square of the speed of sound $\chi(\rho,p) = \frac{\gamma
p}{\rho + p}$, where $\gamma$ is the {\em adiabatic index}, characterizes
a CIG evolution (direct problem), and we obtain all the
thermodynamic quantities of the CIG in terms of the hydrodynamic
quantities $\rho$ and $p$ (specific inverse problem). Moreover, for each of the values of
$\gamma$, we obtain the domain of the variable $\pi = p/\rho$ where the
thermodynamic quantities are positive and the relativistic
compressibility conditions hold. We show that this domain is
relevant for $\gamma >1$.

Section \ref{sec-inverseproblem} is devoted to solving the generic
inverse problem for a CIG indicatrix function, and we show that the
fluids that have the same speed of sound that a CIG are, and
only are, the {\em Poisson gases}, that is, the gases fulfilling
the {\em Poisson law} $p = \beta(s) n^{\gamma}$. Then, we study the compressibility
conditions for the Poisson gases and we analyze the stronger restrictions on the adiabatic index obtained by Taub \cite{Taub} from a kinetic theory approach.

We have pointed out in \cite{CFS-LTE} that energy tensor that
ful\-fill a barotropic relation $p = p(\rho)$ can model particular
evolutions of a non-barotropic perfect fluid. In section
\ref{sec-barotrop} we obtain the barotropic relation of a CIG in
isothermal evolution and we model an isothermal atmosphere and a
self-gravitating isothermal sphere.

The above models have been stablished without reference to any heat equation. Other\-wise, the current relativistic Fourier equations impose a
strong constraint on the temperature when the conductivity does not vanish. In section \ref{sec-esferes}
we take into account this fact in dealing with a model for a
self-gravitating CIG sphere in thermal equilibrium.

In section \ref{sec-s-constant} we obtain the barotropic relation of
a CIG in isentropic evolution and we consider the FLRW models that
fulfill this constraint. The study of the field equations shows that
these models are defined by  a pressure that is a power of the expansion
factor.

In section \ref{sec-solutions} we present preliminary results on
perfect fluid solutions of the field equations that can be
interpreted as a CIG in local thermal equilibrium. For the
case of fluids in irrotational motion we show that our
hydrodynamic characterization implies that, in comoving coordinates,
the pressure is, up to an arbitrary function of the spatial
coordinates, a power of the determinant of the spatial metric. Then, for the spherically symmetric metrics with geodesic motion and 2-sphere curvature changing with the comoving radial coordinate, we show that the only CIG solutions are the FLRW models considered in section \ref{sec-s-constant}.

Interestingly, work in progress seems to show that the known perfect fluid exact solutions to Einstein equations do not verify the CIG constraint exactly. Thus, it can be
suitable to look for exact solutions that approximate a CIG at low
temperatures, which is the range where the CIG model is realistic.
In section \ref{sec-solutions} we also propose a way to approximate a CIG
from a generic ideal gas by using the indicatrix function
$\chi(\rho,p)$.

Finally, in section \ref{sec-remarks} we present several remarks and
report some work in progress.


\section{Local thermal equilibrium: basic concepts and hydrodynamic approach}
\label{sec-lte}

The energetic description of the evolution of a perfect fluid is
given by its energy tensor $T$:
\begin{equation}
\label{perfect-energy-tensor} T = (\rho+ p) u \otimes u + p \, g \,
.
\end{equation}
where $\rho$, $p$ and $u$ are, respectively,  the {\em energy
density}, {\em pressure} and {\em unit velocity} of the fluid.
A divergence-free $T$, $\nabla \cdot T = 0$, of this form is called a
{\em perfect energy tensor}. These conservation equations take the
expression:
\begin{eqnarray}
d p  + \dot{p} u + (\rho + p) a = 0 \, ,  \label{con-eq1} \\[2mm]
\dot{\rho} + (\rho+ p) \theta = 0 \, ,   \label{con-eq2}
\end{eqnarray}
where $a$ and $\theta$ are, respectively, the acceleration and the
expansion of $u$, and a dot denotes the directional
derivative, with respect to $u$, of a quantity $q$, $\dot{q} = u(q)
= u^{\alpha} \partial_{\alpha} q$.

A {\em barotropic evolution} is an evolution along which the {\em
barotropic relation} $ \dif \rho \wedge \dif p = 0$ is fulfilled. It is to be noted that barotropic evolutions may be followed by any fluid, whatever its equations of state, barotropic or not. A perfect
energy tensor describing energetically a barotropic evolution is
called a {\em barotropic perfect energy tensor}.

The thermodynamic scheme for a relativistic perfect fluid is
obtained as the Pascalian and of vanishing heat conductivity restriction of the general
thermodynamic scheme by Eckart \cite{Eckart}. The energy density
$\rho$ is decomposed in terms of the {\em matter density} $n$ and
the {\em specific internal energy} $\epsilon$:
\begin{equation} \label{ro-r-epsilon}
\rho= n(1+\epsilon)  \label{masa-energia} \, ,
\end{equation}
requiring the conservation of matter:
\begin{equation}
\nabla \cdot (nu) = \dot{n} + n \theta = 0 \, .  \label{c-masa}
\end{equation}
When $n=n(\rho,p)>0$, and according to a classical argument, it is
always possible to identify an integral divisor of the one-form
$\Lambda \equiv  (1/n) \dif  \rho + (\rho+p) \dif (1/n)$ with the (absolute)
{\em temperature} $\Theta$ of the fluid, allowing to define the {\em
specific entropy} $s$ by the {\em local thermal equilibrium
equation}:
\begin{equation}
\Theta \dif s = \dif  \epsilon + p \dif (1/n) = (1/n) \dif  \rho + (\rho+p) \dif (1/n)
\, .   \label{re-termo}
\end{equation}
When, in addition to (\ref{con-eq1}) and (\ref{con-eq2}),  a perfect
energy tensor $T$ also satisfies (\ref{masa-energia}),
(\ref{c-masa}) and (\ref{re-termo}), we will say that $T$ evolves in
{\em local thermal equilibrium} (l.t.e.).

We have already shown \cite{Coll-Ferrando-termo}  \cite{CFS-LTE}
that the notion of l.t.e admits a purely hydrodynamic formulation:
{\em a perfect energy tensor} $T$ {\em describes a thermodynamic perfect fluid in l.t.e if, and only
if, its hydrodynamic quantities $\{u, \rho, p\}$ fulfill the hydrodynamic sonic condition:}
\begin{equation}
(\dot{\rho} \dif  \dot{p} - \dot{p} \dif  \dot{\rho}) \wedge \dif \rho \wedge \dif p
= 0     \, .       \label{h-lte}
\end{equation}
When the perfect energy tensor is non isoenergetic,
$\dot{\rho}\not=0$, condition (\ref{h-lte}) states that the
space-time function $\chi \equiv \dot{p}/\dot{\rho}$, called {\em
indicatrix of local thermal equilibrium}, depends only on the
quantities $p$ and $\rho$, $\chi = \chi (p,\rho )$. Moreover this
function of state represents physically the square of the {\em speed
of sound} in the fluid, $\chi (\rho ,p) \equiv  c^2_{s}$.

Note that the above quoted result solves the {\em generic
direct problem}, i.e. the determination of the set ${\bf T}_{\bf F}$ of all the perfect energy tensors
corresponding to all possible energetic evolutions of all perfect fluids ${\bf F}$. In \cite{CFS-LTE} we have also solved the {\em
inverse problem} for a perfect energy tensor $T$, and we have shown
that the set ${\bf F}_T$ of all perfect fluids for which $T$ is the
energetic description of a particular evolution is determined up to
two arbitrary functions of the entropy.

In practice, solving a {\em restricted direct problem} may be more
interesting than solving the generic one. In this way we have solved
in \cite{CFS-LTE} the direct problem for the family of ideal gases
${\bf G}$, which is defined by the equation of state:
\begin{equation}
p = k n \Theta  \, , \qquad \quad    k \equiv {k_B \over m} \,  .
\label{gas-ideal}
\end{equation}
Now the hydrodynamic sonic condition states \cite{CFS-LTE}: {\em
the necessary and sufficient condition for a non barotropic} ($d
\rho \wedge dp \not=0$) {\em and non isoenergetic} ($\dot{\rho} \not
=0$) {\em perfect energy tensor} $T=(u,\rho,p)$ {\em to represent
the l.t.e. evolution of an ideal gas is that the indicatrix function
$\chi\equiv \dot{p}/\dot{\rho}$ be a function of the quantity} $\pi
\equiv p/\rho$, $\chi=\chi(\pi) \not= \pi$.

This statement characterizes the set ${\bf T}_{\bf G}$ of all
the perfect energy tensors which are the energetic evolution of a
non barotropic ideal gas $f \in {\bf G}$. In \cite{CFS-LTE}
we have solved the {\em specific inverse problem} by obtaining, in
terms of $\rho$ and $p$, all the thermodynamic quantities that define
the thermodynamic scheme of an ideal gas.


\section{Classical ideal gas: hydrodynamic characterization}
\label{sec-CIG}

Let us study now the evolution (direct problem) of the {\em
classical ideal gas} (CIG), that is to say, we consider any ideal gas
verifying (\ref{gas-ideal}) and
\begin{equation}
\epsilon = c_v \Theta  \, ,    \label{e-t-cig}
\end{equation}
$c_v > 0$ being the {\em heat capacity at constant volume}. From
(\ref{gas-ideal}) and (\ref{e-t-cig}) it follows that a CIG
satisfies the {\em classical $\gamma$-law}:
\begin{equation}
p=(\gamma-1)n \epsilon  , \qquad \gamma \equiv  1 + {k \over
c_v}   \, ,   \label{g-law}
\end{equation}
$\gamma$ being the {\em adiabatic index}. On the other hand
the l.t.e. equation (\ref{re-termo}) implies that a CIG has the
characteristic equation
\begin{equation}
\epsilon = \epsilon (n, s) =  \bar{\beta}(s) \, n^{\gamma -1}  ,
\quad \bar{\beta}(s) \equiv \exp{\frac{s-\bar{s}_0}{c_v}}  \, .
\label{epsilon-r-s}
\end{equation}
And from (\ref{g-law}) and ({\ref{epsilon-r-s}) it follows that any
CIG fulfills a {\em Poisson law}:\footnote{The classical
$\gamma$-law (\ref{g-law}) and the Poisson law (\ref{poisson}) are
equations of state. A fluid that fulfills (\ref{g-law}) or (\ref{poisson}) is called
a {\em $\gamma$-gas} or a {\em Poisson gas}, respectively. A
Poisson gas in isentropic evolution fulfills a {\em Poisson
adiabatic law} $p = \beta_0  n^{\gamma}$. In section \ref{sec-inverseproblem} Table \ref{table} gives the equations of state that define the diverse families of perfect fluids concerned in this paper and summarizes some of our results.}
\begin{equation}
p = \beta (s) n^{\gamma}   , \quad \beta(s) = (\gamma-1)
\bar{\beta}(s)    \, .               \label{poisson}
\end{equation}

We know \cite{CFS-LTE} that the only intrinsically barotropic ideal
gases are those satisfying $\epsilon(\Theta) = c_v \Theta -1$. Thus
a CIG is, necessarily, non barotropic $d \rho \wedge d p \not=0$,
and we can take the hydrodynamic quantities $(\rho, p)$ as
coordinates in the thermodynamic plane. Then, from the above
expressions (\ref{ro-r-epsilon}), (\ref{gas-ideal}),
(\ref{e-t-cig}), (\ref{g-law}) and (\ref{epsilon-r-s}), we obtain:
\begin{eqnarray}
n(\rho,p) = \rho - {p \over \gamma -1}   \, , \quad 
\epsilon(\rho, p) = {p \over (\gamma -1) \rho - p}    \, , \quad \label{r-e-cig} \\[2mm]
s(\rho,p) = s_0 + c_v \ln{p \over [\rho(\gamma -1) -p]^{\gamma}} \,
.  \quad  \label{s-cig}
\end{eqnarray}
Moreover, the square of the speed of sound can be calculated as
$\chi(\rho,p)= -s_{\rho}' / s_{p}'$, and we obtain:
\begin{equation}
c_s^2 = \chi(\rho,p) \equiv  {\gamma p \over \rho+p}  \, .
\label{chi-cig}
\end{equation}
We summarize these results in the following.
\begin{lemma}  \label{lemma-gic}
In terms of the hydrodynamic quantities $(\rho,p)$, the matter
density $n$, the specific internal energy $\epsilon$, the specific
entropy $s$ and the speed of the sound $c_s$ of a classical ideal
gas are given by {\em (\ref{r-e-cig}), (\ref{s-cig})} and {\em
(\ref{chi-cig})}.
\end{lemma}

As shown in \cite{CFS-LTE}, a non barotropic and isoenergetic
($\dot{\rho} =0$) perfect energy tensor evolves in l.t.e. if, and
only if, it is isobaric, $\dot{p}=0$, and then it admits any
thermodynamic scheme. Thus, for the case of a CIG we have:
\begin{proposition} \label{propo-iso}
The necessary and sufficient condition for a non barotropic and
isoenergetic ($\dot{\rho} =0$) perfect energy tensor $T=(u,\rho,p)$
to represent the l.t.e. evolution of a CIG is that it be
isobaroenergetic: $\dot{\rho}=0$, $\dot{p}=0$. Then $T$ represents
the evolution in l.t.e. of any CIG, and the specific internal energy
$\epsilon$, the matter density $n$, the specific entropy $s$ and the
speed of sound $c_s$ are given by {\em (\ref{r-e-cig}),
(\ref{s-cig})} and {\em (\ref{chi-cig})}.
\end{proposition}
Note that the richness of CIG associated with a non barotropic and
isobaroenergetic perfect energy tensor depends on two parameters,
the adiabatic index $\gamma$ and the heat capacity at constant
volume $c_v$. This last one determines through (\ref{gas-ideal}) and
(\ref{g-law}) the particle mass $m$.

On the other hand, if the perfect energy tensor $T$ is non
isoenergetic, $\dot{\rho} \not= 0$, then the indicatrix function
$\chi \equiv \dot{p}/\dot{\rho}$ equals the square of the speed of
sound \cite{CFS-LTE} and it is given by (\ref{chi-cig}) for a CIG. Conversely, when the perfect
fluid has this indicatrix, the thermodynamic quantities $n$,
$\epsilon$ and $s$ given in (\ref{r-e-cig}) and (\ref{s-cig})
fulfill the CIG characteristic equation (\ref{epsilon-r-s}). Thus we
can state:
\begin{theorem} \label{theorem-CIG}
The necessary and sufficient condition for a non barotropic and non
isoenergetic perfect energy tensor $T=(u,\rho,p)$ to represent the
l.t.e. evolution of a classical ideal gas with adiabatic index
$\gamma$ is that the indicatrix function $\chi\equiv
\dot{p}/\dot{\rho}$ be of the form:
\begin{equation}
\chi = {\gamma p\over \rho + p}   \, .     \label{chi-clas}
\end{equation}
Then, the matter density $n$ and the specific entropy $s$ are given
by {\em (\ref{r-e-cig})} and {\em (\ref{s-cig})}, and the constants
$k$ and $c_v$ are related by
\begin{equation} \label{gamma}
\gamma = 1 + {k \over c_v}  \, .
\end{equation}
\end{theorem}
After this theorem it is worth remarking the following points:
\begin{itemize}
\item[i)]
Theorem \ref{theorem-CIG} offers a purely hydrodynamic
characterization of the set ${\bf C}$ of the CIG by solving the
associated direct problem: the determination of the set ${\bf T}_f$
of all the perfect energy tensors that are the energetic evolution
of any CIG $f \in {\bf C}$. It also solves the {\em specific inverse
problem} for ${\bf C}$: the determination of the set ${\bf C}_T$ of
the CIG whose evolution is described by a given $T \in {\bf T}_{\bf
C}$.
\item[ii)]
In practice, it can be more convenient to verify directly the
condition that characterizes the full set ${\bf C}$ without
specifying the adiabatic index $\gamma$:
\begin{equation} \label{dgamma}
 \dif  {(\rho + p)\dot p \over p\dot \rho } = 0 \, .
\end{equation}
If (\ref{dgamma}) holds, a constant $\gamma$ exists that satisfies
(\ref{chi-clas}), and then the perfect energy tensor $T=(u,\rho,p)$
represents the l.t.e. evolution of a CIG with adiabatic index
$\gamma$.
\item[iii)] Note that in the non isoenergetic case the adiabatic index $\gamma$, that is, the quotient $k/c_v$, is fixed by the hydrodynamic quantities. Remember that  the values $\gamma = 5/3$ and $\gamma = 7/5 $ correspond, respectively, to {\em
monoatomic} and {\em diatomic} ideal gases. Again, the freedom to
choose $c_v$ is related with the particle mass. Some conditions for physical reality imposing constraints on the adiabatic index $\gamma$ will be analyzed below.
\end{itemize}


\subsection{Constraints for physical reality: compressibility conditions}
\label{subsec-cc}

The Pleba\'nski \cite{Plebanski} energy conditions are basic
constraints for physical reality of formal arbitrary media. They impose
algebraic restrictions on the hydrodynamic quantities, and for a
perfect energy tensor take the expression $-\rho < p \leq \rho$. For a medium that fulfills the equation of state of a generic ideal gas (\ref{gas-ideal}), it is reasonable to consider a positive pressure, $p > 0$. Thus, the energy conditions E for these media state:
\begin{equation} \label{e-c}
{\rm E:} \qquad \qquad 0 < \pi \leq 1 \, , \qquad  \pi = \frac{p}{\rho} \, . \qquad \qquad 
\end{equation}

When the perfect fluid evolves in l.t.e. some physical requirements must be
imposed on the thermodynamic quantities. First, the positivity conditions P for the the matter density $n$
and the specific internal energy $\epsilon$: 
\be \label{pos}
{\rm P:} \qquad \qquad  \qquad n>0 \, ,  \qquad \epsilon>0 \, , \qquad \qquad
\ee
and second, the relativistic compressibility
conditions H must be fulfilled \cite{Israel} \cite{Lichnero-1}. We have shown in \cite{CFS-CC} that, for a generic ideal gas, the conditions H can be stated in terms of the indicatrix function, $\chi = \chi(\pi) \not=\pi$:
\begin{equation}
\begin{array}{ll}
 & \qquad  \frac{\pi}{2\pi+1} \!< \! \chi < 1 \,  , \\
{\rm H:} \qquad &   \\
& \qquad  (1+\pi)(\chi-\pi) \chi'  + 2 \chi(1-\chi) >
0 \,   .   
\end{array}   \label{cc-ideal}
\end{equation}
In this paper, where a macroscopic approach is carried out, the conditions E, P and H are called the {\em constraints for physical reality}.

For a CIG, the positivity constraint P can be stated in terms of hydrodynamic quantities as a
consequence of (\ref{r-e-cig}):
\begin{lemma}  \label{lemma-e-r>0}
For a CIG, the the positivity constraint {\rm P}, $n>0$, $\epsilon>0$, hold if,
and only if,
\begin{equation} \label{r-e>0CIG}
\gamma >  1 \, , \qquad  \pi <   {\pi}_m \equiv  \gamma -1 \, .
\end{equation}
\end{lemma}
Now we study the compressibility constraints H for a CIG, that is, after
(\ref{chi-clas}), for
\begin{equation}
\chi(\pi) = \frac{\gamma \pi}{\pi+1} \, , \qquad
\chi'(\pi) = \frac{\gamma}{(\pi+1)^2}   \, . \quad
\label{chi-pi-CIG}
\end{equation}
If $b(\pi)= \frac{\pi}{2 \pi +1}$, we have that both derivatives
$\chi'(\pi)$ and $b'(\pi)$ are positive and decreasing in the
domain $ [0,1]$. Moreover $b(0) = \chi(0) = 0$; $b(1) = 1/3$,
$\chi(1) = \gamma/2$; $b'(0) = 1$, $\chi'(0) = \gamma$.
Consequently, when $\gamma >1$ we have $b(\pi) < \chi(\pi)$ in
the domain $]0,1]$. On the other hand, $\chi < 1$ if $\pi \in
[0, \frac{1}{\gamma-1}[$. Thus, we can state:
\begin{lemma}  \label{lemma-cc-1}
If $\gamma >1$, the first of the compressibility conditions {\rm H} in {\em (\ref{cc-ideal})} holds in the interval:
\begin{equation} \label{cc1-lemma}
0 < \pi < \tilde{\pi}_m \equiv \frac{1}{\gamma-1} \, .
\end{equation}
\end{lemma}

The second compressibility condition in (\ref{cc-ideal}) can be
analyzed taking into account (\ref{chi-pi-CIG}) and we obtain:
\begin{lemma}  \label{lemma-cc-2}
If $\gamma >1$, the second of the compressibility conditions {\rm H} in {\em (\ref{cc-ideal})} holds in the interval:
\begin{equation} \label{cc-2-lemma}
0 < \pi <   \hat{\pi}_m \equiv \   \frac{\gamma + 1}{2\gamma-1} \, .
\end{equation}
\end{lemma}

Finally, note that if $\gamma = 2$, $\pi_m = \hat{\pi}_m = \tilde{\pi}$; if $\gamma \leq 2$, $\pi_m < \hat{\pi}_m < \tilde{\pi}$; and if
$\gamma \geq 2$, $\tilde{\pi}_m < \hat{\pi}_m < \pi_m$. Thus, we can collect
the results of the three lemmas \ref{lemma-e-r>0}, \ref{lemma-cc-1}
and \ref{lemma-cc-2} in the following.
\begin{proposition}  \label{propo-cc-CIG}
For a CIG with adiabatic index $\gamma$, the constraints for physical reality {\rm E}, {\rm P} and {\rm H} are satisfied for values of $\pi$ in a non-empty
subinterval of $[0,1]$ if, and only if, $\gamma >1$. In addition,
they hold in the interval:
\begin{equation} \label{cc-CIG}
\cases{0 < \pi <   {\pi}_m \equiv  \gamma -1 \, , \quad {\rm if} \ \ 1< \gamma \leq 2 \, , \\[3mm] \cr
\displaystyle 0 < \pi <   \tilde{\pi}_m \equiv  \frac{1}{\gamma -1}
\, , \quad {\rm if} \ \  \gamma \geq 2 \, . }
\end{equation}
\end{proposition}
Usually (see, for example \cite{Anile}) the adiabatic index $\gamma$ is supposed to be constrained by 
$1 < \gamma  \leq 2$. Proposition above shows that,
under reasonable physical requirements, the upper limit for $\gamma$
can be relaxed. In subsection \ref{subsec-Taub} we will analyze this apparent contradiction and we will also comment about the stronger restrictions imposed by a model based on the relativistic kinetic theory.


\section{The generic inverse problem for an indicatrix of a classical ideal gas}
\label{sec-inverseproblem}

In Section \ref{sec-CIG} we have solved the specific inverse problem
for the CIG, that is, we have determined the set ${\bf C}_T$ of CIG
having any given perfect energy tensor $T \in {\bf T}_{\bf C}$ as energetic
description. Now we analyze the generic inverse problem for the CIG:
to determine the set ${\bf F}_T$ of all perfect fluids having a given perfect 
energy tensor $T \in {\bf T}_{\bf C}$ as energetic description. The
answer to this problem is tantamount to obtaining the thermodynamics compatible with the expression (\ref{chi-cig}) of the square
of the speed of sound. We can make use of the following known result
\cite{CFS-LTE}:
\begin{lemma}  \label{lemma-inverseproblem}
Let $T$ be a non barotropic and non isoenergetic perfect energy
tensor that evolves in l.t.e. The admissible thermodynamics
are defined by a matter density $n=n(\rho,p)$ and a specific entropy
$s(\rho,p)$ of the form $n=\bar{n}R(\bar{s})$ and $s=s(\bar{s})$,
where $R(\bar{s})$ and $s(\bar{s})$ are arbitrary real functions,
and $\bar{n}(\rho,p)$ and $\bar{s}(\rho,p)$ are, respectively,
particular solutions of the equations:
\begin{equation}
n_{\rho}'  + \chi(\rho,p) n_{p}' = \frac{n}{\rho+p} ,  \qquad
s_{\rho}'  + \chi(\rho,p) s_{p}' = 0    ,   \label{r-s-chi}
\end{equation}
$\chi(\rho,p)$ being the indicatrix function, $\chi \equiv
\dot{p}/\dot{\rho}$.
\end{lemma}
For a CIG the indicatrix function takes the form (\ref{chi-cig}),
with $\gamma >1$, and then the functions $n$ and $s$ given in
(\ref{r-e-cig}) and (\ref{s-cig}) are particular solutions of
(\ref{r-s-chi}). Then, from lemma \ref{lemma-inverseproblem} a
straightforward calculation leads to:
\begin{lemma} \label{lemma-chi-poisson}
If a perfect fluid evolves in l.t.e. and the speed of sound is the
function of state $c_s^2 = \chi(\rho,p)= \frac{\gamma p}{\rho
+p}$, $\gamma>1$, then it is a Poisson gas, that is, it fulfills the
Poisson law:
\begin{equation}
p = \beta (s) n^{\gamma} \, , \qquad  \gamma > 1   \, .
\label{poisson-b}
\end{equation}
\end{lemma}

In order to prove the converse statement we need to know the
characteristic equation for a Poisson gas, which easy follows from
(\ref{poisson-b}) and the local thermal equilibrium relation
(\ref{re-termo}):
\begin{lemma} \label{lemma-ca-eq-poisson}
The thermodynamic characteristic equation of a Poisson gas is given
by:
\begin{equation}
\epsilon(n,s) = \tilde{\beta}(s)n^{\gamma-1} + \alpha(s) \, , \qquad
\tilde{\beta}(s) = \frac{\beta(s)}{\gamma -1} \, .
\label{eq-ca-poisson}
\end{equation}
In particular, when $\alpha (s)=0$ this equation characterizes the {\em
$\gamma$-gases}, that is, the gases fulfilling a classical
$\gamma$-law:
\begin{equation}
p=(\gamma-1)n \epsilon  \, , \qquad \gamma > 1 \, .
\label{g-ley}
\end{equation}
And if, in addition, $\beta(s)= \beta_0 e^s$ is imposed, the CIG is obtained.
\end{lemma}
Note that the characteristic equation (\ref{eq-ca-poisson}) of a
Poisson gas depends on two arbitrary functions of the specific
entropy, $\tilde{\beta}(s)$ and $\alpha(s)$. They correspond to the
two arbitrary functions that give the richness of admissible
thermodynamic schemes of the generic inverse problem presented in
Lemma \ref{lemma-inverseproblem}. Thus the generic specific entropy
is a function of the specific entropy of a CIG given in
(\ref{s-cig}):
\begin{equation}
s(\rho,p) = f(x) \, , \qquad x \equiv  [\rho(\gamma -1) -p]\,
p^{-\frac{1}{\gamma}}  \, .  \label{s-poisson}
\end{equation}
From this expression, the square of the speed of sound can be
calculated as $c_s^2= \chi(\rho,p)= -s_{\rho}' / s_{p}'$, and
we obtain (\ref{chi-cig}). Thus, the indicatrix function of a
Poisson gas is that of a classical ideal gas. This result and Lemma
\ref{lemma-chi-poisson} allow us to state:
\begin{proposition} \label{propo-poisson-iff}
The Poisson gases defined by {\em (\ref{poisson-b})} are, and only
are, the media that evolve in l.t.e. with an indicatrix function of
the form $\chi(\rho,p)= \frac{\gamma p}{\rho +p}$, $\gamma>1$.
\end{proposition}
From Lemma \ref{lemma-ca-eq-poisson} and Proposition
\ref{propo-poisson-iff} it results that the indicatrix function of a
material medium satisfying a classical $\gamma$-law is
(\ref{chi-clas}). Now $\alpha(s) = 0$ corresponds to taking in Lemma
\ref{lemma-inverseproblem} $R(\bar{s})=1$ and $\bar{n}(\rho,p)$
equals the matter density of a CIG. Therefore:
\begin{proposition} \label{propo-gamma-gases}
The $\gamma$-gases, fulfilling a $\gamma$-law, are, and only are, the
media that evolve in l.t.e. with an indicatrix function of the form
{\em (\ref{chi-clas})} and that have a CIG matter
density.
\end{proposition}
%


\begin{table*}

\begin{ruledtabular}
\begin{tabular}{lllll}
   & {\rm Definition} &  {\rm Characteristic Equation \ } & {\rm Equivalent Conditions} &$ \chi =c^2_s \equiv - \frac{s^{\prime}_{\rho}}{s^{\prime}_{p}}$
  \phantom{\Large $\frac{A}{B}$} 
  \\[2mm]
   \hline
{\rm Ideal\ Gas}   \phantom{\Large $\frac{A}{B}$} & $p   = k n
\Theta$ & $\epsilon = \epsilon (s + k \ln n)$ &
 & $\chi = \chi(\pi) \neq \pi$ \\[1mm] \hline
 {\rm Poisson   Gas} \phantom{$E^\frac{E^2}{E}$}  & $p  = \beta(s) n^{\gamma}$ & $\epsilon =
 \tilde{\beta}(s) n^{\gamma -1} + \alpha(s) $& $\chi = \frac{\gamma  \pi}{1+\pi} $  & $\chi = \frac{\gamma
 \pi}{1+\pi} $\\[1mm] \hline

$\gamma-{\rm Gas} $  \phantom{\Large $\frac{A}{B}$}  &$ p =
(\gamma -1) n \epsilon$  & $\epsilon =
 \tilde{\beta} (s) n^{\gamma-1}$  & & $\chi = \frac{\gamma
 \pi}{1+\pi}$ \\[1mm] \hline

 Poisson  Ideal Gas   &
$ p  = k n \Theta , \ \  p  = \beta(s) n^{\gamma}    $ & $ \epsilon =
e^{\frac{s-s_0}{c_v}} n^{\gamma -1} + \epsilon_0 $ & $  p = k n
\Theta ,
 \ \   \epsilon(\Theta) = c_v \Theta + \epsilon_0  \ $ &  $\chi = \frac{\gamma
 \pi}{1+\pi} $ \phantom{\large $E^\frac{E^2}{E_P}$} \\[1.5mm] \hline

 Classical  Ideal Gas \  &  $
p  = k n \Theta, \ \     \epsilon(\Theta) = c_v \Theta \  $ &
   $\epsilon =
e^{\frac{s-s_0}{c_v}} n^{\gamma -1}  $ & $  p = k n \Theta, \ \ p=
(\gamma-1) n \epsilon $ & $ \chi = \frac{\gamma
 \pi}{1+\pi} $ \phantom{\large $E^\frac{E^2}{E_P}$} \\[1mm]

\end{tabular}
\end{ruledtabular}
\caption{Different families of fluids that are considered in this paper (first column) and the functions of state that define them (second column). The third column shows their characteristic equation $\epsilon= \epsilon(s,n)$, and the fourth one shows, for some cases, equations of state that also characterize them. Finally, the fifth column shows the square of the speed of sound $\chi$ in terms of the hydrodynamic quantity $\pi = p/\rho$. Note that each classical ideal gas is a $\gamma$-gas, and each $\gamma$-gas is a Poisson gas, and the three families have the same indicatrix function $\chi(\pi)$. Moreover and ideal gas is a Poisson gas if, and only if, the specific internal energy $\epsilon$ is a linear function of the temperature $\Theta$. And the classical ideal gases are the ideal gases that are $\gamma$-gases.}
\label{table}
\end{table*}

\subsection{Constraints for physical reality}
\label{subsec-Taub}

From the above results, the study of the constraints for physical reality
presented in subsection \ref{subsec-cc} for a CIG can be extended to the
$\gamma$-gases and the Poisson gases. The $\gamma$-gases are submitted
to identical constraints as the CIG as a consequence of Proposition
\ref{propo-gamma-gases}. For the Poisson gases the expressions of the
matter density $n$ and the specific internal energy $\epsilon$ are
more generic than in the case of a CIG. Thus, the second of the positivity constraint P in (\ref{r-e>0CIG}) does not hold necessarily, and then the second compressibility constraint H given by (\ref{cc-2-lemma}) applies for $1< \gamma \leq 2$.
Therefore:
\begin{proposition} \label{propo-cc-poisson}
For a $\gamma$-gas, defined by the condition {\em (\ref{g-ley})}, the
constraints for physical reality are satisfied in the interval given by
{\em (\ref{cc-CIG})}:
$$
\cases{0 < \pi <   {\pi}_m \equiv  \gamma -1 \, , \quad {\rm if} \ \ 1< \gamma \leq 2 \, , \\[3mm] \cr
\displaystyle 0 < \pi <   \tilde{\pi}_m \equiv  \frac{1}{\gamma -1}
\, , \quad {\rm if} \ \  \gamma \geq 2 \, . }. 
$$

For a Poisson gas, defined by the condition {\em (\ref{poisson-b})}, the
constraints for physical reality are satisfied in the interval given
by
\begin{equation} \label{cc-poisson}
\cases{0 < \pi <    \hat{\pi}_m \equiv \   \frac{\gamma + 1}{2\gamma-1} \, , \quad {\rm if} \ \ 1< \gamma \leq 2 \, , \\[1mm] \cr
\displaystyle 0 < \pi <   \tilde{\pi}_m \equiv  \frac{1}{\gamma -1}
\, , \quad {\rm if} \ \  \gamma \geq 2 \, . }
\end{equation}
\end{proposition}
%
%



Most of the thermodynamic expressions presented in previous sections
on CIG and Poisson gases can be found in the literature or can be
deduced from known thermodynamic relations. In particular, from
(\ref{r-e-cig}) and (\ref{chi-cig}) we obtain $c_s^2 = \frac{\gamma
\epsilon(\gamma-1)}{1 + \gamma \epsilon}$, an expression for the
speed of sound that can be found, for example, in \cite{Rezzolla}.
It is also known \cite{Anile} that the speed of sound in a Poisson
gas can be written as (\ref{chi-cig}). Nevertheless, our approach
brings new insights on the subject:
\begin{itemize}
\item[i)]
This approach offers the sufficient condition of this last statement, that is,
it shows that the equation of state (\ref{chi-cig}) characterizes the
Poisson gases.
\item[ii)]
It also emphasizes the purely hydrodynamic nature of this
characterization, and it presents the answer to the inverse problem
for the indicatrix function (\ref{chi-cig}) for the set ${\bf C}$ of  CIG,
the set $\Gamma$ of the $\gamma$-gases, and the set ${\bf P}$ of the
Poisson gases. Moreover, it easily clarifies the inclusion relationship
between these three sets of fluids, ${\bf C} \subset  \Gamma \subset {\bf P}$,
an issue often unclear in the literature, and furthermore it shows that the set ${\bf G} \cap {\bf P}$ of the ideal
gases that are Poisson gases are those with $\epsilon = c_v \Theta + \epsilon_0$; the case $\epsilon_0 =0$ (CIG) arises for a $\gamma$-gas, that is, ${\bf G} \cap \Gamma = {\bf C}$. On the other hand, this approach allows to solve the inverse problem for the energy tensor, leading, in particular, to the result that the set ${\bf P}$ of the Poisson gases are the answer to the
inverse problem for both the CIG and the $\gamma$-gases, that is,
${\bf P} = {\bf F}_{{\bf T}_{\bf C}} = {\bf F}_{{\bf T}_{\bf \Gamma}}$, where ${\bf F}_{{\bf T}_{\bf C}}$ and ${\bf F}_{{\bf T}_{\bf \Gamma}}$ are, respectively, the set of all perfect fluids $f \in {\bf F}$ whose all possible evolution energy tensors ${\bf T}_{\bf C}$ are those corresponding  to any of  the CIG, and the set of all perfect fluids $f \in {\bf F}$ whose all possible evolution energy tensors ${\bf T}_{\bf \Gamma}$ are those corresponding  to any of  the $\gamma$-gases. Table \ref{table} summarizes all these results.
\item[iii)]
Compressibility conditions for a Poisson gas were analyzed in
\cite{Anile}, and sufficient conditions in the quantities $n, p$ were
also presented. Here our approach offers in propositions \ref{propo-cc-CIG} and
\ref{propo-cc-poisson} necessary and sufficient conditions for
the selected constraints for physical reality, and allow to state them in terms of the purely
hydrodynamic equation of state $c_s^2 = \chi(\rho,p)$.
\end{itemize}

Usually, in the literature on thermodynamic perfect fluids  \cite{Anile} \cite{Rezzolla}, the adiabatic index is considered constrained by $1 < \gamma \leq 2$. Nevertheless, our results in propositions \ref{propo-cc-CIG} and \ref{propo-cc-poisson} show that values of the adiabatic index greater than 2 could model physically reasonable
media (CIG, $\gamma$-gases, or Poisson gases) in physically relevant ranges of the hydrodynamic quantity $\pi$. 

The constraint $\gamma \leq 2$ was deduced by Taub \cite{Taub} by imposing the limit $c_s < 1$ for the speed of the sound, a requirement that is included in the compressibility conditions (\ref{cc-ideal}). The contradiction between our statements and the Taub's result is only apparent. Indeed, he deduced the constraint for $\gamma$ by analyzing the behavior of $c_s$ at high temperature, a fact that is consistent with our upper-limit $\pi < \tilde{\pi}$ for $\gamma \geq 2$. 

The difficulties in finding perfect fluid solutions of the Einstein equation modeling realistic media are well known (see, for example, section \ref{sec-solutions}). Thus, in the framework of the General Relativity theory it may be of interest to deal with equations of state that, not being valid in the whole range, are meaningful in a relevant range of the hydrodynamic quantities $\rho$ and $p$. 

In this paper our point of view on the physical constraints is directly macroscopic, not derived from the kinetic theory of gases.  Nevertheless, we believe that it is worthwhile to comment about the Taub inequality obtained from a kinetic approach \cite{Taub}:
\be \label{Taub}
n^2 \leq \rho ( \rho - 3p) \, .
\ee
For a CIG and a $\gamma$-gas the mater density takes the expression $n=n(\rho,p)$ given in (\ref{r-e-cig}). Then, for $\gamma \geq 5/3$ there is no value of $\pi \in ]0,1[$ fulfilling the Taub inequality (\ref{Taub}). Thus, the kinetic theory implies $\gamma < 5/3$ for these media. Nevertheless, the matter density of a Poisson gas is of the form $R(s)n(\rho,p)$, where $R(s)$ is an arbitrary function of the specific entropy. Consequently, for any $\gamma > 1$, we can select in the inverse problem a matter density $\bar{n} = R_0 n(\rho,p)$, with $R_0$ such that (\ref{Taub}) is fulfilled. Moreover, the temperature $\Theta$ can be taken such that the gas ideal equation of state (\ref{gas-ideal}) is verified. Therefore, Poisson ideal gases with  any $\gamma  > 1$ are compatible with the Taub inequality (\ref{Taub}).


\section{Isothermal evolution of a classical ideal gas}
\label{sec-barotrop}

We know \cite{CFS-LTE} that a general (non barotropic) perfect
fluid admits barotropic evolutions. For example, this occurs when
the fluid evolves by keeping constant a determined function of state. When this function of state is not  the specific entropy
$s$, this evolution is, necessarily, isobaroenergetic, $\dot{\rho} =
\dot{p}=0$  \cite{CFS-LTE}. Here we study the isothermal evolution
of a CIG, and we offer two elementary examples.

A generic ideal gas in isothermal evolution fulfills a barotropic
evolution relation\footnote{Not to be confused with an equation of state!} of the form $p = \pi_0 \rho$ \cite{CFS-LTE}. Now we
determine what the constant $\pi_0$ means in the case of a CIG.

If a CIG evolves at constant temperature $\Theta_0$, from
(\ref{ro-r-epsilon}), (\ref{gas-ideal}) and (\ref{e-t-cig}) we
obtain:
\begin{eqnarray}  \label{t-constant}
\frac{p}{\rho} = \pi_0 \equiv  \frac{k \Theta_0}{1+ c_v \Theta_0} =
\frac{(\gamma - 1) \epsilon}{1+\epsilon} < 1 \, , \\   k \Theta_0 =
\frac{(\gamma-1) - \pi_0}{(\gamma-1)  \pi_0} \, . \label{t-constant1}
\end{eqnarray}
Then, we have:
\begin{proposition} \label{prop-t-constant}
A perfect energy tensor $T=(u,\rho,p)$ represents the isothermal
evolution of a CIG if, and only if, it is isobaroenergetic,
$\dot{\rho} = \dot{p}=0$, and the following barotropic relation
holds:
\begin{equation}
p = \pi_0 \rho \, , \qquad  0 < \pi_0 < 1   \, .
\label{t-constant-p-rho}
\end{equation}
Conversely, an isobaroenergetic and barotropic energy tensor with
barotropic relation {\em (\ref{t-constant-p-rho})} represents the
isothermal evolution of any CIG. For a given specific adiabatic
index $\gamma$, the product $k \Theta_0$ is constrained by the
condition {\em (\ref{t-constant1})}
and the specific internal energy $\epsilon$, the matter density $n$,
and the specific entropy $s$ are given by {\em (\ref{r-e-cig})} and
{\em (\ref{s-cig})}, the constants $k$ and $c_v$ being related by
{\em (\ref{gamma})}.
\end{proposition}
%


\subsection{Relativistic model of isothermal atmosphere}
\label{subsec-T-constant-prova}

The external gravitational field to a spherically symmetric object
is given by the Schwarszchild metric:
\be \label{Schw} \dif  s^2 = - \alpha^2 \dif  t^2 + \alpha^{-2} \dif  r^2 + r^2
 \dif  \Omega^2  , \ \  \alpha  \equiv \sqrt{1-\frac{2
\mu}{r}}  . \ee
Let us consider a test classical ideal gas at rest around this
object in a spherical configuration. We have then $\dot{\rho}=
\dot{p}=0$. Moreover, the hydrodynamic equations (\ref{con-eq1}-\ref{con-eq2}) for the unit velocity $u=\alpha^{-1}\partial_t$, the
energy density $\rho( r)$ and the pressure $p( r)$ become:
\be \label{hidro-static} p'( r) = - (\rho+p) \frac{\alpha'}{\alpha}
\, . \ee

Then, as a consequence of proposition \ref{prop-t-constant}, if the
CIG evolves at constant temperature $\Theta_0$, then $p = \pi_0
\rho$, where $\pi_0$ is given in (\ref{t-constant}). Thus,
(\ref{hidro-static}) becomes:
\be \label{hidro-static-t} p'( r) = - 2 \nu p \frac{\alpha'}{\alpha}
\, , \qquad \nu \equiv \frac{1+ \pi_0}{2 \pi_0} \, , \ee
and integrating this equation we have:
\be \label{p-atmosfera-rel} p ( r) = \frac{C}{\alpha^{2 \nu}} =
\frac{C}{\left(1-\frac{2 \mu}{r}\right)^{\nu}} = p_0
\left[\frac{1-\frac{2 \mu}{r_0}}{1-\frac{2 \mu}{r}}\right]^{\nu}   .
\ee

It is worth remarking that the Newtonian model of isothermal
atmosphere of a classical ideal gas leads to:
\be \label{p-atmosfera-clas} p ( r) =  C e^{\frac{\lambda}{r}} = p_0
e^{\lambda \left(\frac{1}{r}- \frac{1}{r_0}\right)} \, , \qquad
\lambda \equiv \frac{\mu}{ k\Theta_0} \, . \ee
For low temperatures, $k \Theta_0 \ll 1$, $\nu$ approaches
$\lambda$, and for weak gravitational field, $\mu \ll r$,
(\ref{p-atmosfera-rel}) and (\ref{p-atmosfera-clas}) have the same
behavior.


\subsection{A model of self-gravitating isothermal sphere}
\label{subsec-T-constant}

The gravitational field generated by a static spherically symmetric
distribution of matter is modeled by the metric:
\begin{eqnarray}
\label{static-sphere} \dif  s^2 = - \alpha^2 \dif  t^2 + \beta^{-2} \dif  r^2 +
r^2 \dif  \Omega^2  , \\ \alpha  = \alpha ( r) , \quad \beta = \beta (
r) \equiv \sqrt{1-\frac{2 m(r)}{r}}  ,
\end{eqnarray}
where the mass function $m(r)$ and the pressure $p(r)$ are submitted
to the differential system:
\begin{eqnarray} \label{chandra-1}
m'( r) = 4 \pi r^2 \rho \, , \\[2mm]
p'(r ) = - \frac{( \rho +p)(m + 4 \pi r^3 p)}{ r ( r-2m)}  \, ,
\label{chandra-2}
\end{eqnarray}
and they fulfill the initial conditions $m(0)=0$ and $p(0) = p_c$.
Moreover, the gravitational potential $\alpha( r)$ can be obtained
from equation (\ref{hidro-static}). A way to close the
Oppenheimer-Volkoff equations (\ref{chandra-1}-\ref{chandra-2}) is
to impose a barotropic constraint $\rho=\rho(p)$.

The solution of the system (\ref{chandra-1}-\ref{chandra-2}) under
the barotropic relation $p = \pi_0 \rho$ was studied by
Chandrasekhar \cite{Chandra}, and he showed that it becomes an
Emden-like equation, as in the classical problem of an isothermal
gas. In this paper by Chandrasekhar and in a more recent one by
Chavanis \cite{Chavanis} this barotropic model is called
relativistic "isothermal" model due to its similarity to the
isothermal Newtonian case.

It is worth remarking that this model is indeed an exact isothermal
relativistic solution as a consequence of proposition
\ref{prop-t-constant}: it performs the evolution at constant
temperature $\Theta_0$ of any ideal gas (we can take any adiabatic
index $\gamma$),   with $\Theta_0$ constrained by
(\ref{t-constant1}). But it could also model any self-gravitating
isothermal generic ideal gas \cite{CFS-LTE} and, particularly, a
Synge relativistic gas \cite{Synge} evolving at constant temperature
as considered in \cite{Thorne}.

Of course, this assertion has been stablished without reference to any 
heat equation and it remains valid under the hypothesis of
vanishing conductivity. Otherwise, the local thermal equilibrium implies
necessarily a gradient of temperature attached to the gradient of
the gravitational potential as a consequence of a result by Tolman
\cite{Tolman-0} \cite{Tolman}. This fact was previously pointed out
in \cite{Thorne} (see also \cite{Chavanis} and next section).


\section{Classical ideal gas spheres in thermal equilibrium}
\label{sec-esferes}

If a fluid has a non-vanishing heat conductivity coefficient $\kappa$,
the energy flux $q$, the temperature $\Theta$ and the fluid
acceleration $a$ are constrained by the relativistic Fourier
equation \cite{Eckart}:
\be \label{Fourier} q = -\kappa (\perp\! \dif  \ln \Theta - a) \, , \ee
where $\perp$ denotes the projector orthogonal to the fluid
velocity.

When the energy flux vanishes we obtain\footnote{It is known that
equation (\ref{Fourier}) leads to a non-causal thermodynamics. The
proposed alternatives in causal extended thermodynamics \cite{Rezzolla} also lead to
$\perp \! \dif  \ln \Theta = a$ when the energy flux vanishes.} $\perp
\! \dif  \ln \Theta = a$ and, if the fluid is at rest with respect the
static gravitational field (\ref{static-sphere}), this condition
leads to:
\be \label{alpha-T} \alpha \Theta = C = constant \, , \ee
and we recover the above-cited result by Tolman \cite{Tolman}.

If we consider a spherical distribution of a CIG in thermal
equilibrium, from (\ref{ro-r-epsilon}), (\ref{gas-ideal}),
(\ref{g-law}) and (\ref{alpha-T}) we obtain:
\be \label{rho-p-alpha} \rho = \left(\frac{1}{kC} \alpha +
\frac{1}{\gamma-1}\right) p \, , \ee
and the hydrostatic equation (\ref{hidro-static}) becomes:
\be p'(\alpha) = - p \left(\frac{1}{kC} +
\frac{\gamma}{\gamma-1}\frac{1}{\alpha}\right)  \, , \ee
which leads to:
\be \label{p-alpha} \displaystyle p(\alpha) =  K \alpha^{
-\frac{\gamma}{\gamma-1}} e^{- \frac{\alpha}{kC}}  \, . \ee
Note that (\ref{rho-p-alpha}) and (\ref{p-alpha}) give a barotropic
relation, $ \dif  \rho \wedge \dif  p = 0$, in a parametric form: $\rho=
\rho(\alpha)$, $p=p(\alpha)$. And from here, we can obtain an implicit barotropic relation $\psi(\rho,p)=0$. On the other hand, from
(\ref{gas-ideal}) and (\ref{alpha-T}) we obtain:
\be \label{r-T} \displaystyle n(\Theta) = \bar{K} \Theta^{
\frac{1}{\gamma-1}} e^{- \frac{1}{k\Theta}}  \, . \ee

If we consider a test classical ideal gas in the Schwarzschild
gravitational field (\ref{Schw}) we have $\alpha^2 = 1- \frac{2
\mu}{r}$. Then (\ref{alpha-T}), (\ref{p-alpha}) and (\ref{r-T}) offer
a model of atmosphere in thermal equilibrium.

Instead, for a self-gravitating distribution, the stelar structure
equations can be written for the functions $m(r )$ and $\alpha( r)$:
\begin{eqnarray} \label{estructura-equilibri-1}
m'( r) = 4 \pi r^2 \rho(\alpha) \, , \\[2mm]
\alpha'(r ) = \alpha \frac{m + 4 \pi r^3 p(\alpha)}{ r ( r-2m)}  \,
,  \label{estructura-equilibri-2}
\end{eqnarray}
where $\rho(\alpha)$ and $p(\alpha)$ are given in
(\ref{rho-p-alpha}) and (\ref{p-alpha}), respectively. A similar
reasoning for the case of a classical monoatomic gas ($\gamma =
5/3$), and in particular the expression (\ref{r-T}) for this
specific case, can be found in the above-cited paper by Tolman
\cite{Tolman-0}.


\section{Isentropic evolution of a classical ideal gas}
\label{sec-s-constant}

When a non barotropic perfect fluid has an isentropic evolution,
this evolution is performed by a barotropic energy tensor, the
barotropic relation $p = \phi(\rho)$ depending on the perfect fluid
characteristic equation \cite{CFS-LTE}. For a CIG, and also for a
$\gamma$-gas or a Poisson gas, an isentropic evolution implies that
the process is polytropic, that is, an adiabatic Poisson law holds:
$p = \beta_0 n^{\gamma}$, $\beta_0= constant$.

From our purely hydrodynamic approach an isentropic evolution means
that the function of state $x(\rho, p)$ given in (\ref{s-poisson})
takes a constant value, and we obtain a specific barotropic
relation. More precisely we have:
\begin{proposition} \label{prop-s-constant}
A perfect energy tensor $T=(u,\rho,p)$ represents the isentropic
evolution of a CIG if, and only if, the following barotropic
relation holds:
\begin{equation}
(\gamma-1) \rho = p + B p^{1/\gamma} \, , \quad B= constant \, .
\label{s-constant-p-rho}
\end{equation}
Conversely, the barotropic evolution {\em (\ref{s-constant-p-rho})}
represents the isentropic evolution of a CIG with adiabatic index
$\gamma$. Moreover, the matter density $n$ is given by
\be \label{s-constant-e-r} n = \frac{B}{\gamma-1}p^{1/\gamma} \, .
\ee
\end{proposition}
The adiabatic Poisson law (\ref{s-constant-e-r}) follows from
(\ref{r-e-cig}) and (\ref{s-constant-p-rho}). The first statement in
proposition above is also valid for both a $\gamma$-gas and a
Poisson gas. And the expression for the matter density in
(\ref{s-constant-e-r}) is valid for a $\gamma$-gas as a consequence
of proposition \ref{propo-gamma-gases}. On the other hand, for a
Poisson gas we evidently have that the matter density fulfills an
adiabatic Poisson law $p = A n^{\gamma}$, where $A$ is now a constant that can be taken independent of $B$.

The analysis of fluids with polytropic evolution has been widely
considered in literature. For example, the study of polytropic
self-gravitating spheres is a basic topic in both Newtonian and
relativistic astrophysics (see for example \cite{Chandra-n}
\cite{Tooper}). A purely hydrodynamic approach to this problem would
imply the study of the relativistic structure equations
(\ref{chandra-1}-\ref{chandra-2}) under the barotropic constraint
(\ref{s-constant-p-rho}). But we do not consider this question here, focusing instead on the analysis of the FLRW universes that model a CIG in
isentropic evolution.


\subsection{Classical ideal gas FLRW models}
\label{sec-FLRW}

The Friedmann-Lema\^itre-Robertson-Walker universes are perfect
fluid space-times with line element:
\be \label{metric-FLRW} \dif  s^2 = - \dif  t^2 + \frac{R^2(t)}{1+ \frac14
\varepsilon r^2}( \dif r^2 + r^2 \dif  \Omega^2)  , \ \ \varepsilon = 0, 1,
-1  , 
\ee
and homogeneous energy density and pressure given by:
\begin{eqnarray} \label{Friedmann-eq-1}
\rho = \frac{3 \dot{R}^2}{R^2}  +  \frac{3 \varepsilon}{R^2} \equiv \rho(R) \, , \\[2mm]
p = - \rho - \frac{R}{3} \partial_{R} \rho \equiv p(R) \, .
\label{Friedmann-eq-2}
\end{eqnarray}
Evidently, we have a barotropic evolution, $ \dif  \rho \wedge \dif  p =0$.
Nevertheless, in looking for physically relevant models we must
impose a physically realistic barotropic relation $p=\phi(\rho)$.
Then, (\ref{Friedmann-eq-2}) enables us to determine $\rho(R)$, and
(\ref{Friedmann-eq-1}) becomes a {\em Friedmann equation} for
$R(t)$. The significant cosmological models for radiation and matter
dominant eras are obtained by taking $\rho = 3p$ and $p=0$,
respectively.

What are the {\em generalized Friedmann equations} when the energy
content is a classical ideal gas with adiabatic index $\gamma$? The
homogeneity of the hydrodynamic quantities $\rho$ and $p$ implies
that, necessarily, the evolution is at constant entropy. Then, as a
consequence of proposition \ref{prop-s-constant}, the barotropic
relation is of the form (\ref{s-constant-p-rho}). A straightforward
calculation allows us to integrate equation (\ref{Friedmann-eq-2})
and to determine $p(R)$, and then $\rho(R)$, and one obtains:
\begin{proposition} \label{prop-FLRW-ideal}
The classical ideal gas FLRW models are defined by the generalized
Friedmann equation {\em (\ref{Friedmann-eq-1})} with the energy
density $\rho(R)$ given by:
\begin{equation}
\rho(R) = n_0 \left(\frac{R_0}{R}\right)^3 + \frac{p_0}{\gamma-1}
\left(\frac{R_0}{R}\right)^{3\gamma}   . \label{FLRW-ideal}
\end{equation}
In terms of the expansion factor $R$, the pressure, the matter
density and the temperature are given by:
\begin{eqnarray} \label{FLRW-ideal-p} p(R) = p_0 \left(\frac{R_0}{R}\right)^{3
\gamma} , \quad n(R) = n_0 \left(\frac{R_0}{R}\right)^{3} , \quad  \\
\label{FLRW-ideal-p1} \Theta(R) = \frac{p_0}{k n_0}
\left(\frac{R_0}{R}\right)^{3 (\gamma - 1)}  .  \qquad \quad 
\end{eqnarray}
\end{proposition}
For $\gamma=5/3$ we obtain a model of monoatomic gas, and for
$\gamma=7/5$ a model of diatomic gas. The cosmological model with
decoupled matter and radiation follows by taking $\gamma=4/3$. And
with a limiting procedure we can recover the pressure-less solution
($p_0=0$) and any $\gamma$-model, $p=(\gamma-1) \rho$, if $n_0=0$,
and in particular the radiation-dominant solution ($\gamma=4/3$).

It is worth remarking that the results in proposition
\ref{prop-FLRW-ideal} also apply to model a $\gamma$-gas in
expansion. And for a Poisson gas the model depends on one more
parameter allowing two different values for the constants $n_0$ in
(\ref{FLRW-ideal}) and in (\ref{FLRW-ideal-p}) and
(\ref{FLRW-ideal-p1}).

The above classical ideal gas FLRW models have been achieved by
imposing the barotropic relation (\ref{s-constant-p-rho}).
Nevertheless, these models can also be obtained by requiring $p
\propto R^{-3 \gamma}$. Indeed, equation (\ref{Friedmann-eq-2}) can
be solved under this assumption and we obtain the expression
(\ref{FLRW-ideal}) for $\rho(R)$. Thus, we have:
\begin{corollary} \label{cor-1}
The classical ideal gas FLRW models in proposition
\ref{prop-FLRW-ideal} are, and only are, the FLRW models with a
pressure of the form {\em (\ref{FLRW-ideal-p})}.
\end{corollary}
It is worth remarking that all the FLRW models with pressure
depending on the metric function $R$ as (\ref{FLRW-ideal-p}) can be
interpreted as a CIG in isentropic evolution.


\section{On the classical ideal gas solutions of Einstein Equations}
\label{sec-solutions}

After theorem \ref{theorem-CIG}, the general form for the CIG field
equations follows by adding constraint (\ref{chi-clas}),
$\dot{p}/\dot{\rho} = {\gamma p\over \rho + p}$, to the usual
perfect fluid field equations. And, according to the energy
conservation condition (\ref{con-eq2}), this constraint is
equivalent to:
\be \label{dotp-theta} \frac{\dot{p}}{p} = - \gamma \, \theta \, .
\ee

In some specific kinematic or thermodynamic situations this
condition is simpler. Thus, in previous sections we have considered
the field equations for a CIG under isothermal or isentropic
evolutions, both cases leading to a barotropic evolution relation.

Let us consider now a CIG solution with irrotational motion. Then,
in comoving coordinates the metric tensor takes the form:
\be \label{metric} \dif  s^2 = - e^{2 \nu} \dif  t^2 + g_{i j} \dif  x^i \dif  x^j
\, . \ee
The fluid expansion is $\theta = (\ln \Delta)^{\cdot}$, where
$\Delta^2$ is the determinant of the spatial metric $g_{i j}$. Then,
from (\ref{dotp-theta}) we obtain:
\be \label{p-Delta} p = P(x^i) \Delta^{-\gamma} \, , \qquad  \Delta
\equiv \sqrt{|g_{i j}|} \, . \ee
Thus, the field equations for an irrotational CIG follow by
replacing the pressure $p$ for the expression (\ref{p-Delta}) in the
field equations in comoving coordinates.

The above result applies to the spherically symmetric metrics and to
their hyperbolic and parabolic counterparts. Now the metric tensor
takes the form:
\be \label{metric} \dif s^2 = - e^{2 \nu(t,r)} \dif  t^2 + e^{2
\lambda(t,r)} \dif  r^2 + Y^2(t,r) \dif  \Omega^2 \, , \ee
where $d \Omega^2$ is a metric of constant curvature $k = 0, 1, -1$.
Then (\ref{p-Delta}) can be written as:
\be \label{p-ssst} p^{-\frac{1}{\gamma}} =  3 h(r) Y^2 e^{\lambda}
\, , \ee
a condition that can be added to the usual set of field equations if
we look for a classical ideal gas model.


\subsection{Solutions in geodesic motion and admitting a G$_3$ on S$_2$ with $Y'(r)\not=0$}
\label{subsec-solutions-G3}

Now we focus on the case $Y' \not=0$ and a geodesic motion, that is,
$\nu=0$. Then, the field equations can be partially integrated and
one has \cite{Kramer}:
\begin{eqnarray} \label{ssst-1}
p(t)  Y^2 = - 2 Y \ddot{Y} - \dot{Y}^2 - \varepsilon f^2(r)  , \quad \varepsilon = 0, 1, -1  , \qquad  \\[2mm]
e^{\lambda} = Y' F(r)  , \quad F^2[k - \varepsilon f^2] = 1  ,
\quad k=0,1,-1  . \qquad  \label{ssst-2}
\end{eqnarray}
Then, the CIG condition (\ref{p-ssst}) leads to
$p^{-\frac{1}{\gamma}} =  3 h(r) F(r) Y^2 Y' = h(r) F(r) (Y^3)'$,
and we obtain:
\begin{eqnarray}
 \label{ssst-3} Z^2 \equiv Y^3 = \alpha(t) a(r) + \beta(t) , \qquad \\
  \label{ssst-31}  
 \alpha(t) \equiv p^{-\frac{1}{\gamma}}  , \qquad  a(r) \equiv \int \frac{d
r}{h(r)F(r)}  \, .
\end{eqnarray}
Note that $Y' \not= 0$ implies $a'(r)\not=0$ and $\alpha(t) \not=0$.
Moreover, when $Y$ factorizes we have a vanishing shear and an homogeneous expansion \cite{Kramer}. Consequently, the FLRW limit occurs when $\beta = c \alpha$, $c = constant$.

On the other hand, the change $Z^2 = Y^3$ allows us to write the
field equation (\ref{ssst-1}) as \cite{Kramer}:
\be \label{ssst-4} \ddot{Z} + \frac34 p(t) Z +\frac34 \epsilon
f^2(r) Z^{-\frac13} = 0 \, .
 \ee
When $\epsilon =0$, equation (\ref{ssst-4}) can be solved by quadratures for each election of the function $p(t)$ \cite{Bonaetall}. But we are interested here in CIG solutions. Then, substituting the expressions (\ref{ssst-3}) and
(\ref{ssst-31}) for $Z$ and $p$ in (\ref{ssst-4}) we have:
\be
A a^2 + B a + C + 3 \varepsilon f^2 (\alpha a + \beta)^{\frac43} = 0  \, , \label{ssst-5} 
\ee  
\begin{eqnarray}
A = A(t) \equiv 2 \alpha \ddot{\alpha} - \dot{\alpha}^2 + 3 \alpha^{2-\gamma} \, ,  \, \ \ \quad \qquad \\[1mm]
B = B(t) \equiv 2 [\alpha \ddot{\beta} + \beta \ddot{\alpha}  - \dot{\alpha} \dot{\beta} + 3 \beta \alpha^{1- \gamma}] \, , \\[1mm]
C = C(t) \equiv 2 \beta \ddot{\beta} - \dot{\beta}^2 + 3 \beta^2
\alpha^{- \gamma}  \, . \quad \qquad  \
\end{eqnarray}
Thus, we have proved:

\begin{lemma}
The metrics {\em (\ref{metric})} with $Y' \not=0$ that model a
classical ideal gas with geodesic motion are defined by four
functions $\alpha(t)\not=0$, $\beta(t)$, $f(r)$ and $a(r)$, $a'(r)
\not=0$, submitted to the differential equation {\em
(\ref{ssst-5})}. The metric functions $Y(t,r)$ and $\lambda(t,r)$
are given in {\em (\ref{ssst-3}-\ref{ssst-31})} and {\em
(\ref{ssst-2})}, respectively, and $\nu=0$.
\end{lemma}

Equation (\ref{ssst-5}) can be written as:
\begin{eqnarray}
M^3 = \varphi \, N^4 , \, \ \qquad \varphi = \varphi(a) \equiv - 27 \varepsilon f^6 \, , \label{ssst-6} \\[1mm]
M = A a^2 + B a + C  \, , \qquad N = \alpha a + \beta \, .  \label{ssst-7}
\end{eqnarray}
If we isolate $\varphi$ in (\ref{ssst-6}) and differentiate with respect $t$ we obtain:
\be
3 \dot{M} N - 4 M \dot{N} = 0 \, ,  \label{ssst-8}
\ee
and, taking into account (\ref{ssst-7}), we arrive to:
\be
P a^3 + Q a^2 + R a + S + 3 = 0  \, , \label{ssst-9} 
\ee  
\begin{eqnarray}
P = P(t) \equiv 3 \alpha \dot{A} - 4 \dot{\alpha} A  \, ,  \, \qquad \qquad \qquad \\[1mm]
Q = Q(t) \equiv 3 \beta \dot{A} + 3 \alpha \dot{B}  - 4 \dot{\alpha} B - 4 \dot{\beta} A \, , \\[1mm]
R = R(t) \equiv 3 \beta \dot{B} + 3 \alpha \dot{C}  - 4 \dot{\beta} B - 4 \dot{\alpha} C \, , \\[1mm]
S = S(t) \equiv 3 \beta \dot{C} - 4 \dot{\beta} C  \, . \, \qquad  \qquad \qquad  \
\end{eqnarray}

If $a'(r) \not=0$, equation (\ref{ssst-9}) implies the four equations
$P(t)=Q(t)=R(t)=S(t)=0$ for the two functions $\alpha(t)$ and $\beta(t)$.
From these equations we obtain:
\be 
0= \frac14 [\alpha^3 S - \beta \alpha^2 R + \beta^2 \alpha Q - \beta^3 P] =
 \alpha^6
\left[\left(\frac{\beta}{\alpha}\right)^{\! .}\right]^3 \!  , 
\ee
and, consequently, $\beta = c \alpha$. Then, the four equations reduce to:
\be 
A = K \alpha^{\frac43} \, , \quad \varphi(a) = K^3(a + c)^2 \, , \quad K = constant \,  . 
\ee
Thus, the metric becomes a FLRW model (\ref{metric-FLRW}) with expansion factor $R(t) = \alpha^{\frac13}$. Moreover, the pressure is $p \propto R^{-3 \gamma}$, and taking into account corollary \ref{cor-1}, we have proved:
\begin{proposition} \label{prop-nosolutions}
The only perfect fluid solutions with geodesic
motion and admitting a G$_3$ on $S_2$ with $Y'(r) \not=0$,
which can be interpreted as classical ideal gases, are the FLRW
models labeled in proposition
{\em \ref{prop-FLRW-ideal}}.
\end{proposition}
%


\subsection{On the exact solutions that approach a classical ideal gas behavior}
\label{subsec-aprox}

The preliminary result given in the previous subsection is an example
of the difficulties in looking for exact solutions to field equations
that model a classical ideal gas in local thermal equilibrium. Work
in progress reveals this fact. Indeed, elsewhere
\cite{CFS-parabolic} we have studied class II Szekeres-Szafron
models in local thermal equilibrium and, although we have found
physically realistic solutions, none of them represents a classical
ideal gas. A similar situation occurs in analyzing the thermodynamic
meaning of the Stephani universes, a task we undertook years ago
\cite{C-F}: there are no exact classical ideal gas models.
Nevertheless, Stephani models representing a generic ideal gas, and
approximating a classical one at low temperatures can be found (see
also the more recent paper \cite{CFS-CC}).

The classical ideal gas is a physically realistic model only at low
temperatures. Consequently, it may be interesting to obtaining
solutions with other thermodynamic schemes but with analogous
behavior at low temperatures. The most natural way to achieve this
is to do what we did in \cite{C-F} (see also\cite{CFS-CC}) for the
Stephani universes: i.e., to look for a generic ideal gas that
approximates a classical one.

The generic ideal gases are characterized by an indicatrix function
$\chi \equiv \dot{p}/\dot{\rho}$ depending only on the hydrodynamic
quantitie $\pi \equiv p/\rho$, $\chi=\chi(\pi) \not= \pi$
\cite{CFS-LTE}. In the particular case of a classical ideal gas this
function and its first derivative are given in (\ref{chi-pi-CIG}).
The k-th derivative is:
\be \chi^{(k)}(\pi) =  \gamma \frac{(-1)^{k+1}  k!}{(1+
\pi)^{k+1}}  , \quad \chi^{(k)}(0) = \gamma (-1)^{k+1}  k!
\, .  \ \ \ee
Consequently,
\be \chi(\pi) = \gamma \, \sum_{k=1}^{\infty} (-1)^{k+1}
\pi^{k} \, . \ee
Then, we say that a generic ideal gas with indicatrix function $\chi
= \chi(\pi)$ approximate a classical one at m-th order if:
\be \chi(\pi) = \gamma \, \sum_{k=1}^{m} (-1)^{k+1}  \pi^{k}
+ O(\pi^{m+1}) \, . \ee
The Stephani ideal gas models obtained in \cite{C-F} admit a
subfamily that approximate at first order classical ideal gas
models. Moreover these models fulfill the compressibility conditions
in a wide range of the interval $[0,1]$. Nevertheless, the class II
Szekeres-Szafron metrics that are ideal gas models only approximate
a classical ideal gas at zero order \cite{CFS-parabolic}.


\section{Ending comments and work in progress}
\label{sec-remarks}

In this paper the hydrodynamic approach to the local thermal
equilibrium developed in \cite{Coll-Ferrando-termo}, 
\cite{CFS-LTE} and \cite{CFS-CC} have been applied to the classical ideal gas. Thus,
the specific direct and inverse problems for the CIG case have been
solved for both isoenergetic (proposition \ref{propo-iso}) and non
isoenergetic evolutions (theorem \ref{theorem-CIG}). The
compressibility conditions for the CIG have been revisited from this
hydrodynamic perspective, and we have studied, for any given
adiabatic index $\gamma >1$, the domain of the hydrodynamic quantity
$\pi = p/\rho$ where they hold (proposition \ref{propo-cc-CIG}). The
analysis of the extended inverse problem for a CIG shows that the
Poisson gases are characterized by having the indicatrix function of
a CIG (proposition \ref{propo-poisson-iff}), and the $\gamma$-gases
are the Poisson gases with a CIG matter density (proposition
\ref{propo-gamma-gases}). The compressibility conditions have also
been analyzed for the $\gamma$-gases and the Poisson gases
(proposition \ref{propo-cc-poisson}).

A barotropic perfect energy tensor can model the evolution of a non
barotropic fluid \cite{CFS-LTE}. Here we obtain the barotropic
relation that a CIG in isothermal evolution fulfills (proposition
\ref{prop-t-constant}), and we apply this result to model an
isothermal atmosphere (subsection \ref{subsec-T-constant-prova}), an
isothermal self-gravitating sphere (subsection
\ref{subsec-T-constant}), and a CIG sphere in thermal equilibrium
(section \ref{sec-esferes}). The barotropic relation of a CIG in
isentropic evolution (proposition \ref{prop-s-constant}) allows us
to present a FLRW solution that models a CIG in l.t.e. (proposition
\ref{prop-FLRW-ideal}).

Further work will be devoted to obtaining perfect fluid solutions
that model CIG in non barotropic evolution. A first result in this
direction presented in section \ref{sec-solutions} shows that the
only CIG spherically symmetric solutions in geodesic motion and $Y'\not=0$ are the above-mentioned FLRW models
(proposition \ref{prop-nosolutions}). A similar limited result seems
to derive from the study of the Szekeres-Szafron solutions that
model a CIG \cite{CFS-parabolic}. One can overcome this situation
in looking for exact solutions that approximate a CIG at low
temperatures. A way to control this approximation has been outlined
in subsection \ref{subsec-aprox}.

In our search for physically realistic solutions of the Einstein
equations we can directly add the CIG hydrodynamic condition
(\ref{dotp-theta}) (or (\ref{chi-clas})) to the common perfect fluid
equations, or we can perform a wider study that includes perfect
fluids and thermodynamic schemes differing from that of a CIG. This
is the method we have built in studying thermodynamic
Szekeres-Szafron solutions \cite{CFS-parabolic}, and likewise the one we shall use for the analysis of the thermodynamic thermodynamic behavior of other families of perfect
fluid solutions hereafter.

For a family of perfect fluid solutions of the Einstein field
equations with perfect energy tensor $T \equiv (u, \rho,p)$, our method consist in the following steps. In a first step
we impose the generic hydrodynamic constraint (\ref{h-lte}) and
obtain the indicatrix function $\chi = \chi(\rho, p)$ for the
subfamily that verifies it. In a second step we detect the subfamily
with an ideal gas indicatrix by imposing $\chi = \chi(\pi)
\not=\pi$. Finally, in a third step, when this function does not coincide with the CIG
indicatrix (\ref{chi-clas}) for any value of the involved
parameters, we can look for solutions that approximates a CIG as
proposed in subsection \ref{subsec-aprox}.


\begin{acknowledgements}
This work has been supported by the Spanish ``Ministerio de
Econom\'{\i}a y Competitividad", MICINN-FEDER project
FIS2015-64552-P.
\end{acknowledgements}


\nocite{*}
\bibliography{CIGPRD}

\end{document}